\documentstyle[12pt]{article}

\begin{document}
\thispagestyle{empty}
{\baselineskip0pt
\leftline{\large\baselineskip16pt\sl\vbox to0pt{\hbox{DAMTP} 
               \hbox{University of Cambridge}\vss}}
\rightline{\large\baselineskip16pt\rm\vbox to20pt{
               \hbox{DAMTP-1998-143}
               \hbox{UTAP-308}
               \hbox{RESCEU-55/98}
               \hbox{\today}
\vss}}%
}
\vskip15mm

\begin{center}
{\large\bf Inverse Cascade of Primordial Magnetic Field in MHD Turbulence}
\end{center}

\begin{center}
{\large Tetsuya Shiromizu
\footnote{JSPS Postdoctal Fellowship for Research Abroad}} \\
\vskip 3mm
\sl{DAMTP, University of Cambridge \\ 
Silver Street, Cambridge CB3 9EW, UK \\
\vskip 5mm
Department of Physics, The University of Tokyo, Tokyo 113-0033, 
Japan \\
and \\
Research Centre for the Early Universe(RESCEU), \\ 
The University of Tokyo, Tokyo 113-0033, Japan
}
\end{center}
\vskip 5mm
\begin{center}
{\it to be published in Physics Letter B}
\end{center}
\begin{abstract}
The feature of the spectrum of primordial magnetic field is
studied by using renormalization group analysis in
magnetohydrodynamics. Taking account of the renormalized resistivity 
at the fixed point, we show that the scaling of the typical scale 
with time obeys $L(t) \sim t^{2/5}$ for random initial condition.   
\end{abstract}


\vskip 1cm

The magnetic fields observed in the various astrophysical scales\cite{Obs} 
is one of the important problems in cosmology\cite{PMF}. The 
generation mechanism of magnetic field in proto-galaxy scale was 
already proposed by Kulsrud et al\cite{Cen}.   
However, as this mechanism cannot explain 
the magnetic fields in intergalactic mediums and interclusters regions 
where matter is poor, it seems that the problem of the origin of 
magnetic fields still remains. This origin 
is often attributed to the very early universe. Some attractive  
mechanisms in the course of cosmological phase transitions
have been proposed \cite{PT}. 
In these mechanisms, the coherent length of the generated 
magnetic field is much smaller than the galaxy scale etc., 
because the magnetic field is generated by micro physical 
processes. 
Thus, one needs to 
investigate the process in which the small magnetic domains 
evolve to larger scales observed in the present universe. Since 
Kulsrud and Anderson\cite{KA} pointed out 
that the kinetic dynamo theory\cite{Dynamo} breaks down in 
interstellar mediums one cannot expect 
enough amplification of the magnetic field. Hence, we must re-analyze 
the evolution of MHD (Magnetohydrodynamics) from the starting 
point. 

Recently, Brandenburg et al\cite{IC1}\cite{IC2} showed numerically  
that the inverse cascade occurs by using cascade model, 
and the occurrence of inverse cascade is also showed 
analytically by Olesen \cite{Olesen}. 
In this paper,  
we reconsider Olesen's study from the viewpoint of the 
renormalization group. We show that the case considered in Ref.
\cite{Olesen} corresponds to our analysis at the fixed point.  
Then we discuss the evolution of the coherent length near the 
fixed point.  

Almost all of astrophysical systems have high magnetic Reynolds number
because it is highly conductive in the typical astrophysical scales.
Thus, MHD turbulence will appear in such system with fluid flow and we
can use the statics of the turbulence to analize this system. We will
apply the result obtained in Ref.\cite{RG} to the evolution of
primordial magnetic field in our study.

First of all, we review the renormalization group analysis of MHD. 
The basic equations of the incompressible MHD are given by  
%
\begin{eqnarray}
\partial_t {\bf P}+{\bf Q}\cdot\nabla {\bf P}=-\nabla p_*
+\gamma_+ \Delta {\bf P}+\gamma_-\Delta {\bf Q}+{\bf f} ,
\end{eqnarray}
%
and
%
\begin{eqnarray}
\partial_t {\bf Q}+{\bf P}\cdot\nabla {\bf Q}=-\nabla p_*
+\gamma_+ \Delta {\bf Q}+\gamma_-\Delta {\bf P}+{\bf g}, 
\end{eqnarray}
%
where ${\bf P}={\bf v}+{\bf B}$, ${\bf Q}={\bf v}-{\bf B}$, 
$p_*=p+(1/2){\bf B}^2$, and $\gamma_\pm =\nu \pm \eta $. 
The ${\bf f}$ and ${\bf g}$ are the 
stirring forces and they are assumed to satisfy the statistical 
correlation,
%
\begin{eqnarray}
\langle f_i(k) f_j(q) \rangle = 2k^{-y}A_0(2\pi)^4 J_{ij}({\bf k})
\delta^4 (k+q)
\end{eqnarray}
%
%
\begin{eqnarray}
\langle f_i(k) g_j(q) \rangle = 2k^{-y}B_0(2\pi)^4J_{ij}
({\bf k})\delta^4 (k+q) ,
\end{eqnarray}
%
and
%
\begin{eqnarray}
\langle g_i(k) g_j(q) \rangle = 2k^{-y}A_0(2\pi)^4J_{ij}({\bf k})
\delta^4 (k+q),
\end{eqnarray}
%
where $J_{ij}({\bf k})=\delta_{ij}-k_ik_j/|{\bf k}^2|$ and 
$A_0, B_0$ are constants. 
The power of $k$ should be decided by the initial condition. 
These forces are introduced so that the above system is equivalent to
the original system in the inertial range({\it correspondence
principle}\cite{YO}). 
The solution is written as
%
\begin{eqnarray}
{\hat G}_0^{-1}{\hat P}_i(k)={\hat f}_i(k)-iJ_{ijk}({\bf k}) 
\int dq \pmatrix{Q_j(k-q) P_k(q) \cr P_j(k-q)Q_k(q) \cr}, 
\end{eqnarray}
%
where ${\hat P}_i(k)=\pmatrix{P_i(k) \cr Q_i(k) \cr}$, 
${\hat f}_i(k)=\pmatrix{f_i(k) \cr g_i(k) \cr}$ and 
$J_{ijk}({\bf k})=k_jJ_{ik}({\bf k})$. 
${\hat G}_0(k)$ is the bare Green function written as
%
\begin{eqnarray}
{\hat G}^{-1}_0(k) = \pmatrix{-i\omega +\gamma_+ |{\bf k}|^2 & 
\gamma_- |{\bf k}|^2  \cr 
\gamma_- |{\bf k}|^2 & -i \omega +\gamma_+ |{\bf k}|^2 \cr},
\end{eqnarray}
%

Now, we divide the $k-$space into two parts, that is, large scale 
modes $0< k < \Lambda e^{-r}$ and small scale modes 
$\Lambda e^{-r} < k < \Lambda$, where $\Lambda$ is the ultraviolet
cut off.  Then, we construct the effective equation for the large 
scale mode by integrating out the small scale modes. 
The result becomes
%
\begin{eqnarray}
{\hat G}^{< -1}(k;r){\hat P}^{<}_i(k)={\hat f}^{<}_i(k)
-i \lambda_0 J_{ijk}({\bf k})
\int dq \pmatrix{Q_j^{<}(k-q) P_k^{<}(q) \cr P_j^<(k-q)Q_k^<(q) \cr}, 
\end{eqnarray}
%
where subscript $<$ means $k \in [0, \Lambda e^{-r}]$, and ${\hat
G}(k;r)$ is the dressed Green function. $\lambda_0$ is the 
expansion parameter introduced technically.  
Now, $\gamma_{\pm}$ depends on
the renormalization parameter, $r$, and obey the following 
equation,
%
\begin{eqnarray}
\frac{d\gamma_{\pm}(r)}{dr}=\frac{\lambda_0^2}{4}A_dA_0\frac{\gamma_+^2(r)}
{\gamma_-^2(r)}\frac{1}{\nu^2(r)\eta^2(r)}
\frac{e^{\epsilon r}}{\Lambda^\epsilon}F_{\pm}(\gamma_+, \gamma_-), 
\end{eqnarray}
%
where $A_d=1/(120\pi^{3/2})$, $\epsilon = 1+y$, and 
$F_{\pm}$ are the functions of $\gamma_\pm$, whose explicit 
expressions are not given here. 
 From the above equation, one finds that the solution at the fixed point is 
$ \gamma_{\pm} \propto e^{\epsilon r/3}$\cite{RG}. 

Compared the eq. (6) with eq. (7), one can see easily that the system is 
invariant under the following scale transformation,
%
\begin{eqnarray}
{\bf k} \rightarrow  {\tilde {\bf k}}={\bf k} e^r =:{\bf k}\ell~~~~~~~~~
\omega \rightarrow  {\tilde \omega}=\omega \ell^\alpha
\end{eqnarray}
%
%
\begin{eqnarray}
{\bf B}(\omega \ell^\alpha , {\bf k}\ell) 
\rightarrow {\tilde {\bf B}}({\tilde {\bf k}}, {\tilde \omega})
={\bf B}({\bf k},\omega) \ell^{-\beta}~~~
{\bf v}(\omega \ell^\alpha , {\bf k}\ell) 
\rightarrow {\tilde {\bf v}}({\tilde {\bf k}}, {\tilde \omega})=
{\bf v}({\bf k}, \omega) \ell^{-\beta}
\end{eqnarray}
%
%
\begin{eqnarray}
\nu (\ell ) \rightarrow {\tilde \nu}=\nu (\ell) \ell^{\alpha -2}~~~
\eta (\ell ) \rightarrow {\tilde \eta}=\eta (\ell) \ell^{\alpha -2}
\end{eqnarray}
%
%
\begin{eqnarray}
{\bf f}({\bf k}, \omega) \rightarrow  {\tilde f}({\tilde {\bf k}}, 
{\tilde \omega})={\bf f}({\bf k}, \omega)\ell^{\alpha -\beta}~~~
{\bf g}({\bf k}, \omega) \rightarrow  {\tilde g}({\tilde {\bf k}}, 
{\tilde \omega})={\bf g}({\bf k}, \omega)\ell^{\alpha -\beta}.
\end{eqnarray}
%
 From the requirement for the invariance of the stirring forces, 
one can obtain the relation, 
%
\begin{eqnarray}
2\beta=3 \alpha+(y+3).
\end{eqnarray}
%

Now we can consider the evolution of the power spectrum $E(k,t)$,
%
\begin{eqnarray}
\langle B^2 \rangle = \int dk E(k,t)
\end{eqnarray}
%
 From the scaling of (10)$\sim$(13), the equation 
%
\begin{eqnarray}
\ell^{5+2\alpha -2 \beta}E(k,t)=E(\ell k, \ell^{-\alpha}t) 
\end{eqnarray}
%
holds. Defining the function $\psi (k,t)=k^{-5-2\alpha-2\beta} E(k,t)$, the 
above equation becomes
%
\begin{eqnarray}
\psi(\ell k, \ell^{-\alpha}t)=\psi (k,t).
\end{eqnarray}
%
The solution is written as
%
\begin{eqnarray}
\psi (k,t)=\psi(k^\alpha t).
\end{eqnarray}
%
Thus, the spectrum is given by 
%
\begin{eqnarray}
E(k,t)=k^{2-\alpha-y}\psi(k^\alpha t)
\end{eqnarray}
%
As $\alpha=\frac{5-y}{3}$ holds at the fixed point\cite{RG}, we obtain 
%
\begin{eqnarray}
E(k,t)=k^{(1-2y)/3}\psi(k^{(5-y)/3} t)
\end{eqnarray}
%
at the fixed point. 
The above expression is the same as one obtained by Olesen\cite{Olesen}. 
Furthermore, as $\partial_t E(k,t) \sim E(k,t)/t \sim 
\eta (k)k^2 E(k,t)$ approximately at the 
fixed point, where $\eta (k) \propto k^{-\epsilon/3}$\cite{RG}, the time 
evolution of the typical length obeys 
%
\begin{eqnarray}
L(t) \sim k^{-1} \propto t^{3/(5-y)}.
\end{eqnarray}
%

Let us consider the case where the initial spectrum is white noise, $y=-5/2$, 
that is, $E(k,0) =k^2 \psi ( 0)$. 
In this case, the typical correlation length becomes 
$L(t) \propto t^{2/5}$. If we assume that the magnetic field is 
generated in a cosmological phase transition at the 
temperature $T_f$(the cosmic time $t_f$) and 
the coherent scale is the same as the bubble size, 
the present physical coherent scale is given by 
%
\begin{eqnarray}
L^{\rm wn}(t_0) & = & L^{\rm wn}(t_f)\frac{a(t_0)}{a(t_f)}
\Bigl(\frac{t_0}{t_f}\Bigr)^{2/5} \nonumber \\
& \simeq & 10^{35}\Bigl( \frac{f_b}{10^{-4}}\Bigr)
\Bigl(\frac{T_f}{100{\rm GeV}}\Bigr)^{-1/5}{\rm GeV}^{-1} 
\simeq 10^{-2} \times 10 {\rm Mpc},
\end{eqnarray}
%
where $10 {\rm Mpc}$ is the Silk damping scale of the magnetic 
field\cite{Silk} where the magnetic field with smaller 
scale is damped by the photon diffusion around the recombination epoch. 
$f_b$ is the ratio of the bubble size $L(t_f)$ to the
horizon scale $H_f^{-1}$. For the electroweak phase transition, 
these typical values becomes $f_b \sim 10^{-4}$ and 
$T_f \sim 100{\rm GeV}$\cite{Ratio}. 
This coherent length is $10^{12}$ times as much as the comovingly 
developed scale; $\sim 10{\rm AU}$. 
We note that this result differs from 
one obtained by numerical simulations in cascade model, 
$L(t) \propto t^{0.25}$\cite{IC1}. 
However, since we used the renormalized resistivity 
which contains the non-linear effect fully, our estimation might be 
more useful.  

Next, to compare with the above result, we consider the case of 
Kraichnan spectrum $E(k,0) \propto k^{-3/2}$, 
that is, $L(t) \propto t^{4/3}$, 
which is the one expected by the naive argument of Alfv\'{e}n effect 
in the inertial range\cite{MHD}. 
In this case, the naive estimation shows that the coherent scale 
exceeds the present horizon scale as follows,
%
\begin{eqnarray}
L^{\rm kra}(t_0) & \simeq & 10^{63}\Bigl( \frac{f_b}{10^{-4}}\Bigr)
\Bigl(\frac{T_f}{100{\rm GeV}}\Bigr)^{8/3}{\rm GeV}^{-1} \nonumber \\
& \simeq & 10^{28} \times 10 {\rm Mpc} \gg H_0^{-1}\sim 3000{\rm Mpc}.
\end{eqnarray}
%
In this estimation, we need to study in the frame work of 
general relativity. 

 From the above investigation, the correlation length can become 
enough size to explain the macroscopic scale. Although the Silk damping effect 
erases the structure 
smaller than the Silk scale, there is no drastic change of the quantitative 
result after taking account of the reasonable viscosity
term\cite{IC2}. 
Finally, we note that these results does not depend on the helicity of
the magnetic field. 

\section*{Acknowledgements}
The author would like to thank Masahiro Morikawa and Ryoichi Nishi 
for their comments. The author is grateful to Gary Gibbons and DAMTP 
relativity group for their hospitality. He also thanks T. Uesugi 
for a careful reading of the manuscript of this paper. This work 
is supported by JSPS fellow.


\begin{thebibliography}{99}

\bibitem{Obs}
P.P. Kronberg, Pep. Prog. Phys. {\bf 57}, 325(1994)
\bibitem{PMF}
For example, A. V. Olinto, `Cosmological Magnetic Field' in the Proceeding 
of 3rd RESCEU Symposium `Particle Cosmology', eds. K. Sato,
T. Yanagida and T. Shiromizu, Univ. Acad. Press, p151(1998);\\
K. Enqvist, astro-ph/9803196
\bibitem{Cen}
R. M. Kulsrud, R. Cen, J. P. Ostriker and D. Ryu, Astrophys. J. {\bf 480}(1997),481;\\
B. D. G. Chandran, Astrophys. J. {\bf 482}(1997),156;\\
B. D. G. Chandran and O. Rodriguez, Astrophys. J. {\bf 490}(1997), 156;\\
B. D. G. Chandran, Astrophys. J. {\bf 492}(1998), 179 
\bibitem{PT}
B. Cheng and A. V. Olinto, Phys. Rev. {\bf D50}(1994),2421;\\
G. Baym, D. B\"{o}deker and L. McLerran, Phys. Rev. {\bf D53}(1996), 662;\\
G. Sigl, A. V. Olinto and K. Jedamik, Phys. Rev. {\bf D55}(1997), 4582;\\
T. Shiromizu, Phys. Rev. {\bf D58}(1998), 107301
\bibitem{KA}
R. M. Kulsrud and S. W. Anderson, Astrophys. J. {\bf 396}(1992)606
\bibitem{Dynamo}
For example, E. N. Parker, Cosmological Magnetic Field (Oxford Univ. 
Press, Oxford, 1979)
\bibitem{IC1}
A. Brandenburg, K. Enqvist and P. Olesen, Phys. Rev. {\bf D54}(1996)1291
\bibitem{IC2}
A. Brandenburg, K. Enqvist and P. Olesen, Phys. Lett. {\bf B391}(1997)395
\bibitem{Olesen}
P. Olesen, Phys. Lett. {\bf B398}(1997)321
\bibitem{RG}
S. J. Camargo and H. Tasso, Phys. Fluids {\bf B4}(1992)1199
\bibitem{YO}
V. Yakhot and S. A. Orszag, J. Sci. Computing, {\bf 1}(1986),3;\\
W. P. Dannevik, V. Yakhot and S. A. Orszag, Phys. Fluids {\bf
30}(1987), 2021
\bibitem{Silk}
K. Jedamzik, V. Katalinic and A. V. Olinto, Phys. Rev. {\bf D57} (1998), 3264
\bibitem{Ratio}
A. F. Heckler, Phys. Rev. {\bf D51}(1995), 405
\bibitem{MHD}D. Biskamp, Nonlinear Magnetohydrodynamics (Cambridge 
Univ. Press, 1993)
\end{thebibliography}
\end{document}